\documentclass[]{spie}  
\usepackage[dvips]{graphicx}

\title{The BaR-SPOrt Experiment}

\author{ M. Zannoni\supit{a,h}, S. Cortiglioni\supit{b}, G. Bernardi\supit{b},
E. Carretti\supit{b}, S. Cecchini\supit{b},  C. Macculi\supit{b},
\\E. Morelli\supit{b}, C. Sbarra\supit{b}, G. Ventura\supit{b}, L.
Nicastro\supit{c}, J. Monari\supit{d}, M. Poloni\supit{d}, S.
Poppi\supit{d}, \\V. Natale\supit{e}, M. Baralis\supit{f}, O.
Peverini\supit{f}, R. Tascone\supit{f}, G. Virone\supit{f}, A.
Boscaleri\supit{g}, E. Pascale\supit{g}, \\G. Boella\supit{h}, S.
Bonometto\supit{h}, M. Gervasi\supit{h}, G. Sironi\supit{h}, M.
Tucci\supit{i}, R. Nesti\supit{j}, R. Fabbri\supit{k}, \\P. de
Bernardis\supit{l}, M. De Petris\supit{l}, S. Masi\supit{l}, M.V.
Sazhin\supit{m}, E.N. Vinyajkin\supit{n} \skiplinehalf
\supit{a}CNR
I.A.S.F.-Milano, \supit{b}CNR I.A.S.F.-Bologna, \supit{c}CNR I.A.S.F.-Palermo, Italy\\
\supit{d}CNR I.R.A.-Bologna, Italy, \supit{e}CNR I.R.A.-Firenze, Italy\\
\supit{f}CNR I.R.I.T.I., Italy\\
\supit{g}CNR I.R.O.E., Italy\\
\supit{h}Univ. di Milano - Bicocca, Milano, Italy\\
\supit{i}Instituto de Fisica de Cantabria, Santander, Spain\\
\supit{j}Osservatorio Astrofisico di Arcetri, Firenze, Italy\\
\supit{k}Univ. di Firenze, Firenze, Italy\\
\supit{l}Univ. di Roma La Sapienza, Roma, Italy\\
\supit{m}Moscow State University-Moscow,
\supit{n}N.I.R.F.I.-Novgorod, Russia}

\authorinfo{Further author information: (Send correspondence to M. Zannoni)\\
M. Zannoni: E-mail: zannoni@mi.iasf.cnr.it, Telephone: +39 02 23699331\\
Address: Istituto di Astrofisica Spaziale e Fisica Cosmica sezione
di Milano, via Bassini 15, I-20133 Milano Italy }

\begin{document}
  \maketitle

\begin{abstract}
BaR-SPOrt (Balloon-borne Radiometers for Sky Polarisation
Observations) is an experiment to measure the linearly polarized
emission of sky patches at 32 and 90 GHz with sub-degree angular
resolution. It is equipped with high sensitivity correlation
polarimeters for simultaneous detection of both the U and Q stokes
parameters of the incident radiation. On-axis telescope is used to
observe angular scales where the expected polarization of the
Cosmic Microwave Background (CMBP) peaks. This project shares most
of the know-how and sophisticated technology developed for the
SPOrt experiment onboard the International Space Station. The
payload is designed to flight onboard long duration stratospheric
balloons both in the Northern and Southern hemispheres where low
foreground emission sky patches are accessible. Due to the
weakness of the expected CMBP signal (in the range of $\mu$K),
much care has been spent to optimize the instrument design with
respect to the systematics generation, observing time efficiency
and long term stability. In this contribution we present the
instrument design, and first tests on some components of the 32
GHz radiometer.
\end{abstract}



\section{INTRODUCTION}
\label{sect:intro}  

The Cosmic Microwave Background (CMB) is the most sensitive tool
to investigate the very high red-shift Universe
\cite{1982PLB....115...189S,1990bmc..book.....D,1996PhRvD..54.1332J,1997ApJ...488....1Z,1999MNRAS.304...75E}.
Since its discovery in 1964 \cite{1965ApJ...142..419P}, positive
and precise results have been collected about the spectrum
\cite{1985ApJ...291L..23S,1996ApJ...473..576F,1999tkc..conf..165Z}
and the spatial distribution
\cite{1996ApJ...464L...1B,2000Natur.404..955D,2000ApJ...545L...5H}.
Current and future space missions like
MAP\footnote{http://map.gsfc.nasa.gov/} and
Planck\footnote{http://astro.estec.esa.nl/Planck/} are mainly
devoted to the all-sky mapping of CMB small-scale anisotropies for
which they will reach the highest sensitivities. A measure of the
degree of residual polarization of the CMB, against its unique
capability to solve the degeneracy among cosmological parameters
that anisotropy alone is not able to remove
\cite{1982PLB....115...189S,1985SvAL...11..204S,1995ApL....32..105S,1997ApJ...488....1Z},
is not available yet. All the attempts to detect it produced only
upper limits (see Table \ref{polupplim}), since either the
foreseen polarized component of the CMB is definitely lower than
the instrumental sensitivities or the contribution of the
systematics dominates the final error budget. The most effective
way to overcome such problems is to design dedicated experiments.
One of them will be the space mission
SPOrt\footnote{http://sport.bo.iasf.cnr.it/}\cite{2002apb..conf...109C}.
BaR-SPOrt, a balloon experiment funded by ASI (Italian Space
Agency), shares most of the know-how and technological development
of the SPOrt mission. While SPOrt has for scientific target a
nearly all sky polarization mapping at 22 and 32 GHz and a
tentative detection of the CMBP at large angular scales
($>7^{\circ}$) at 60 and 90 GHz, the goal of BaR-SPOrt is to
detect the CMBP in some low foreground regions with sub-degree
angular resolution.

\begin{table}[h]
\begin{center}
\begin{tabular}{lllll}
\hline {\bf Resolution (deg)} & {\bf Frequency~(GHz)} & {\bf Sky
Coverage} &
{\bf Upper Limit} & {\bf Reference} \\
\hline
$15$            & $4$       & Scattered                                 & 300 mK                & \cite{1965ApJ...142..419P} \\
$1.5-40$        & $100-600$ & $GC$                                      & 3-0.3 mK               & \cite{1978PhRvD..17.1901C} \\
$15$            & $9.3$     & $\delta = +40^\circ$                      & 1.8 mK                & \cite{1979ApJ...232..341N} \\
$15$            & $33$      & $+37^\circ \leq \delta \leq +63^\circ$    & 180 $\mu$K            & \cite{1981ApJ...245....1L} \\
$18''-160''$    & $5$       & $\delta = +80^\circ$                      & 4.2 mK - 120 $\mu$K   & \cite{1988Natur.331..146P} \\
$1.2$           & $26-36$   & $NCP$                                     & 30 $\mu$K             & \cite{1993ApJ...419L..49W} \\
$1.4$           & $26-36$   & $NCP$                                     & 18 $\mu$K             & \cite{1995ApJ...445L..69N} \\
$7$             & $33$      & $SCP$                                     & 267 $\mu$K            & \cite{1997NewA....3....1S} \\
$6'$            & $8.7$     & $\delta=-50^\circ$                        & 16 $\mu$K             & \cite{2000MNRAS.315..808S} \\
$0.24$          & $90$      & $NCP$                                     & 11 $\mu$K             & \cite{2002ApJ...573L..73H} \\
$7$             & $26-36$   & $\delta = +43^\circ$                     & 10 $\mu$K             & \cite{2001ApJ...560L...1K} \\
\hline
\end{tabular}
\end{center}
\caption{Existing upper limits for the CMB linear polarization.}
\label{polupplim}
\end{table}

\section{Scientific Motivations and Targets}
\begin{figure} [h]
\begin{center}
\begin{tabular} {c}
\includegraphics[width=8.5cm]{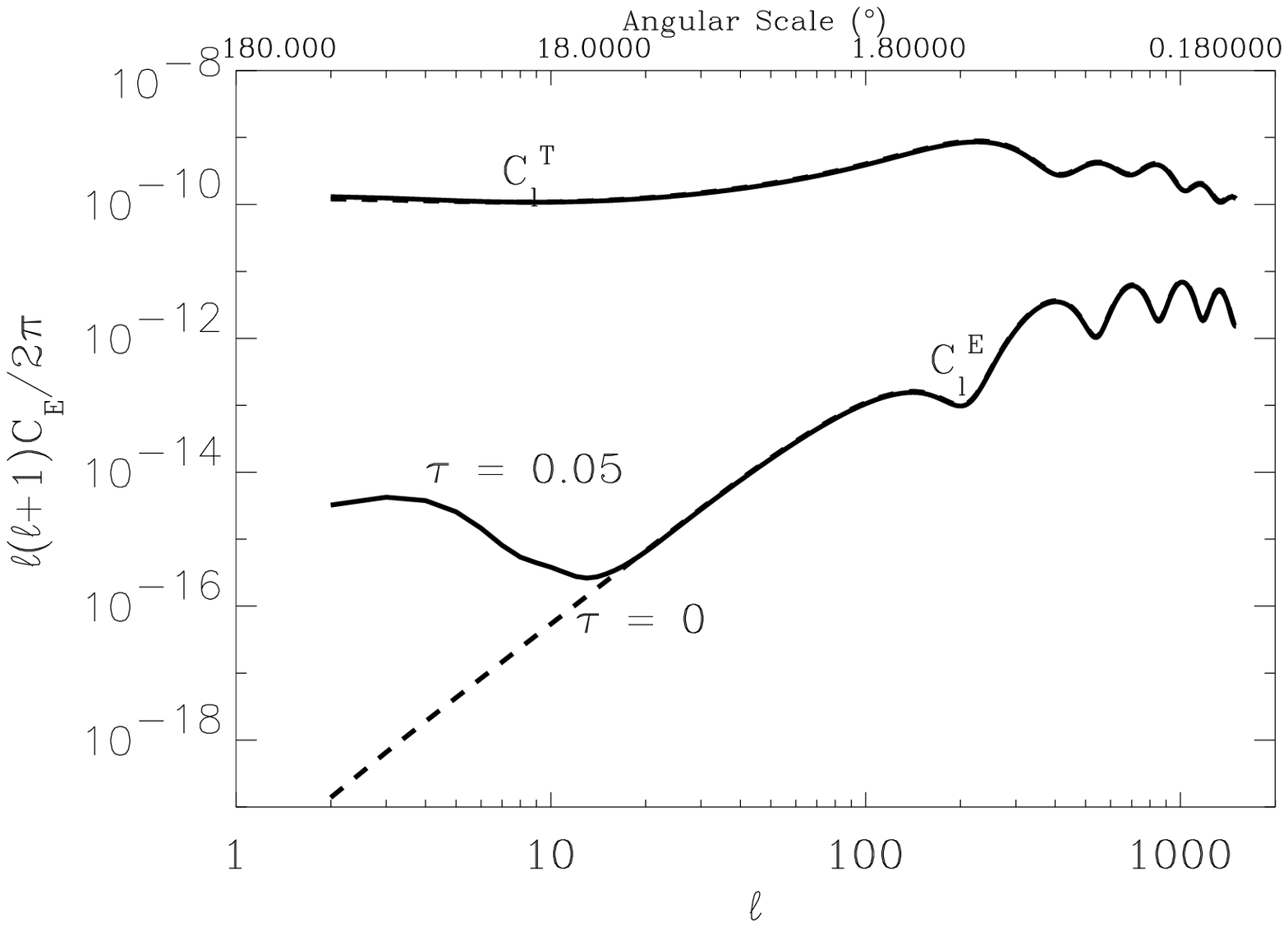}
\hspace{0.01 cm}
\includegraphics[width=8.5cm]{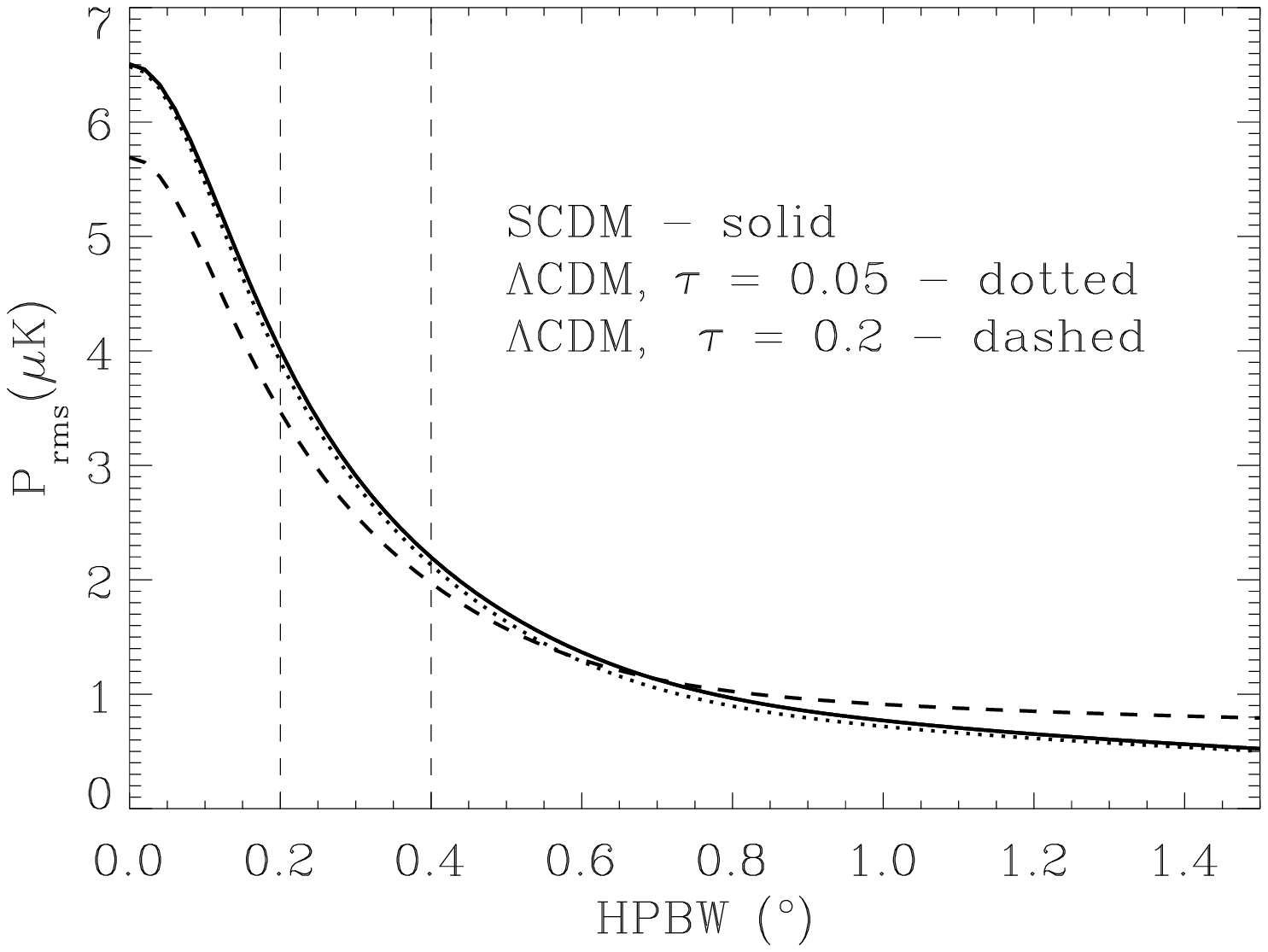}
\end{tabular}
\end{center}
\caption{{\em Left:} anisotropy and E-mode power spectra. The
expected polarized component of the CMB peaks in the sub-degree
($l \geq 100$) scales while the larger (degree) angular scales are
more sensitive to the reionization scenarios. \newline {\em
Right:} the sky $P_{rms}=(<Q^{2}>+<U^{2}>)^{1/2}$ versus the beam
width for different cosmological models. Vertical lines point out
the BaR-SPOrt HPBWs at 32 and 90 GHz} \label{polspectra}
\end{figure}

Detection of the extremely low signal level expected for the large
scale polarization of the CMB (see Figure \ref{polspectra}) needs
a very stable environment. This is the reason why all-sky surveys
can be only carried out in space where long observing time are
feasible and very quiet and stable conditions exist. Ground-based
and balloon-borne experiments can give scientific results
observing small sky patches where they can reach sensitivities
comparable with the expected level of polarization. As pointed out
in Ref.~\citenum{2001NewA....6..173C}, ground-based instruments
operating in the millimeter domain are plagued by atmospheric
emission which is the main source of spurious polarizations even
in the best observing site like the Antarctic plateau. The
instrumental polarization can correlate the unpolarized
atmospheric signal producing offset, whose fluctuations can then
degrade the expected sensitivity. A possible way out is to perform
the observations from stratospheric altitude where the residual
atmosphere contribution and its fluctuations are lower. The
limited observing time (a long duration flight lasts for about two
weeks - cfr \cite{2000Natur.404..955D} but longer are feasible)
reduces the typical targets to small sky patches ($\sim4^\circ
\times 4^\circ$ and smaller). On such portions of sky, sensitive
polarization measurements at sub-degree scales, where the expected
polarization peaks (see Figure \ref{polspectra} {\em left panel}),
are reachable goals for instruments like BaR-SPOrt, for which new
state of the art millimeter devices have been developed. As shown
in Figure \ref{polspectra}, the expected {\em rms} polarization,
$P_{rms}$, is maximum at small angular scales and is only weakly
dependent on cosmological models. BaR-SPOrt in the 90 GHz
configuration, due to the beam-size of $0.2^\circ$, has the
capability to detect CMBP (see Table \ref{maincharacteristics})
independently of the cosmological model. In the 32 GHz
configuration, where the beam size is $0.4^\circ$ ($P_{rms} \sim
2.1 \mu$K), BaR-SPOrt will be able to improve the current upper
limit on CMBP.

\begin{figure} [bht]
\includegraphics[width=5.4cm]{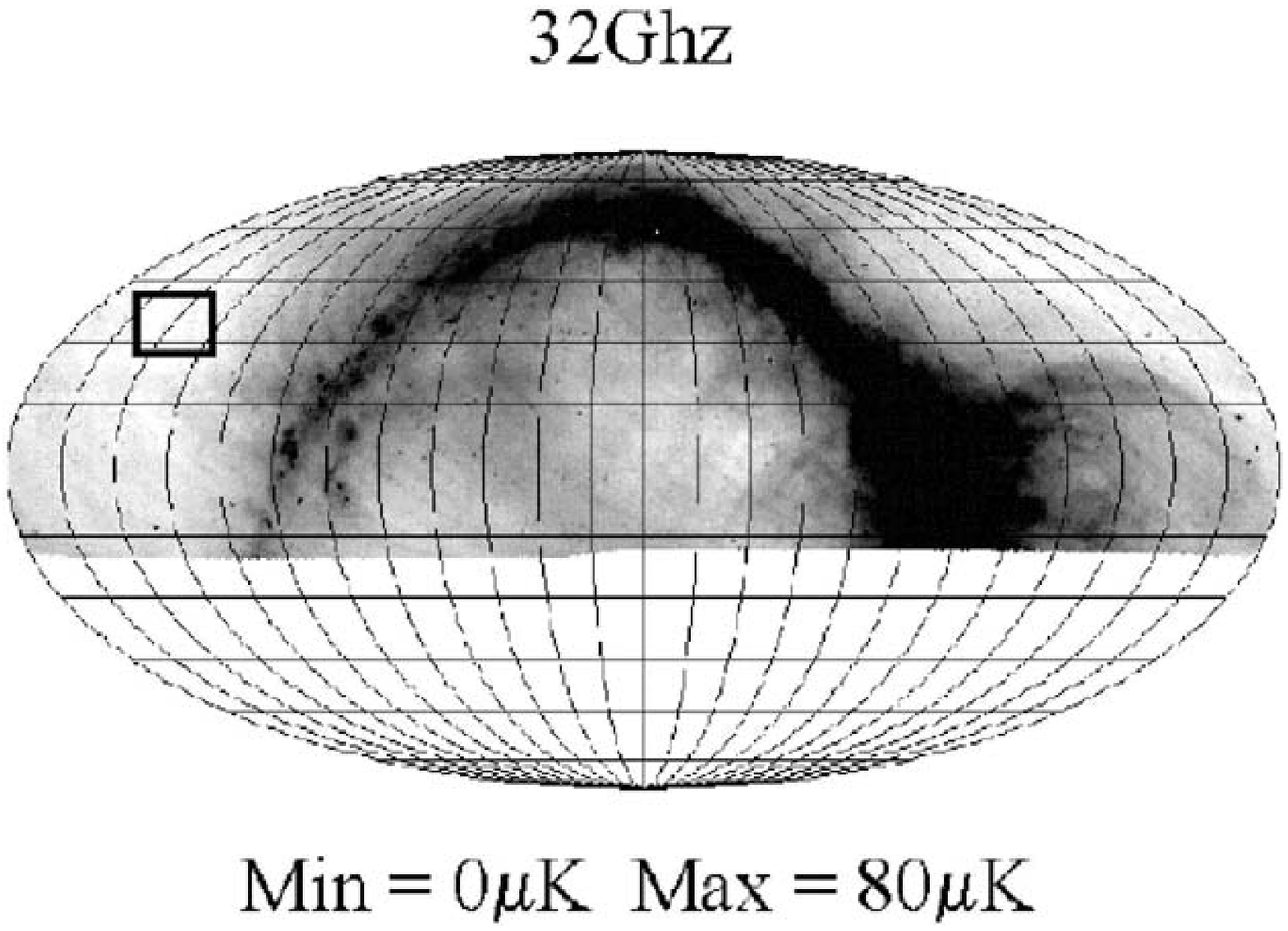}
\hspace{0.01 cm}
\includegraphics[width=5.4cm]{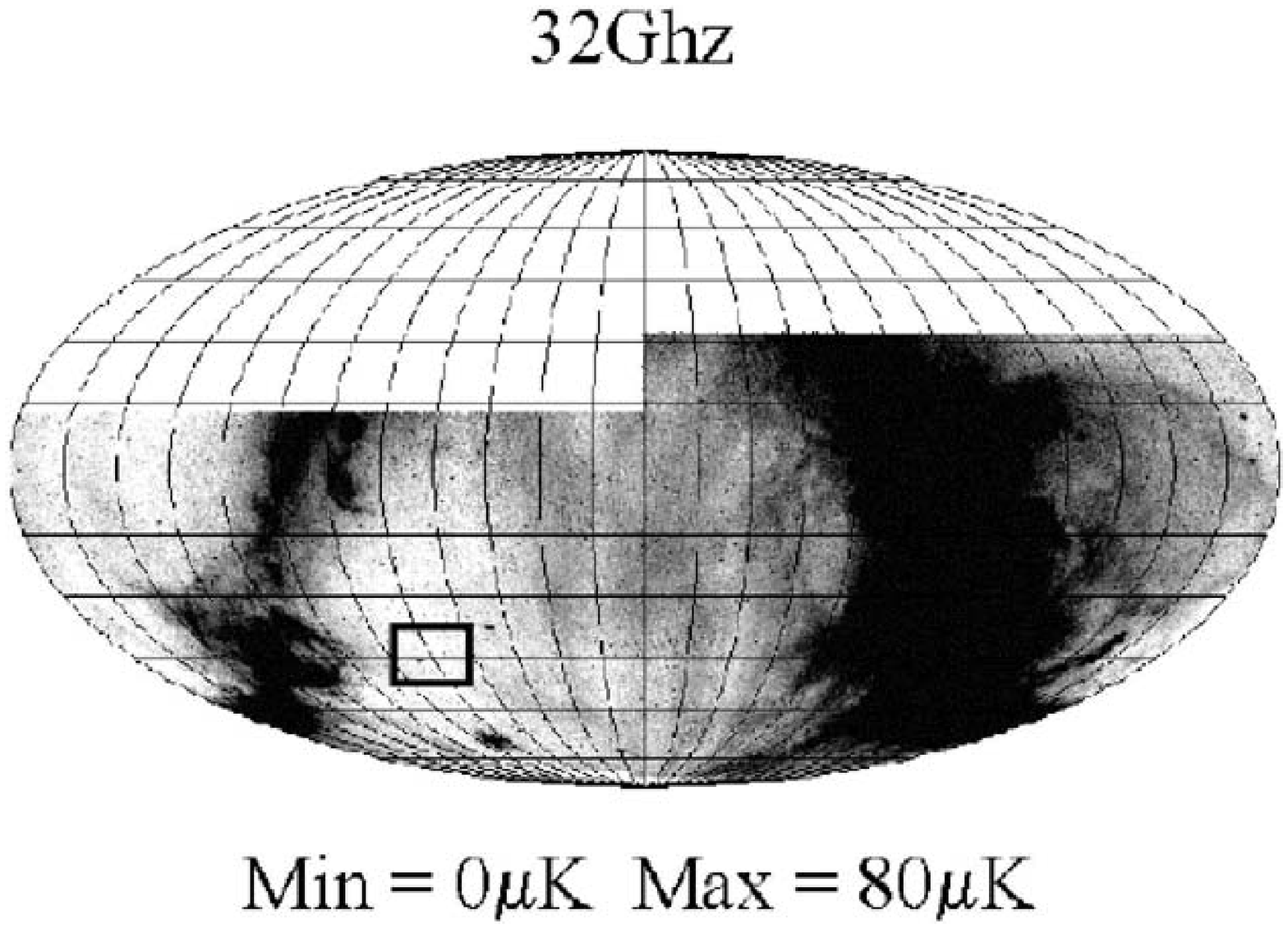}
\hspace{0.01 cm}
\includegraphics[width=5.4cm]{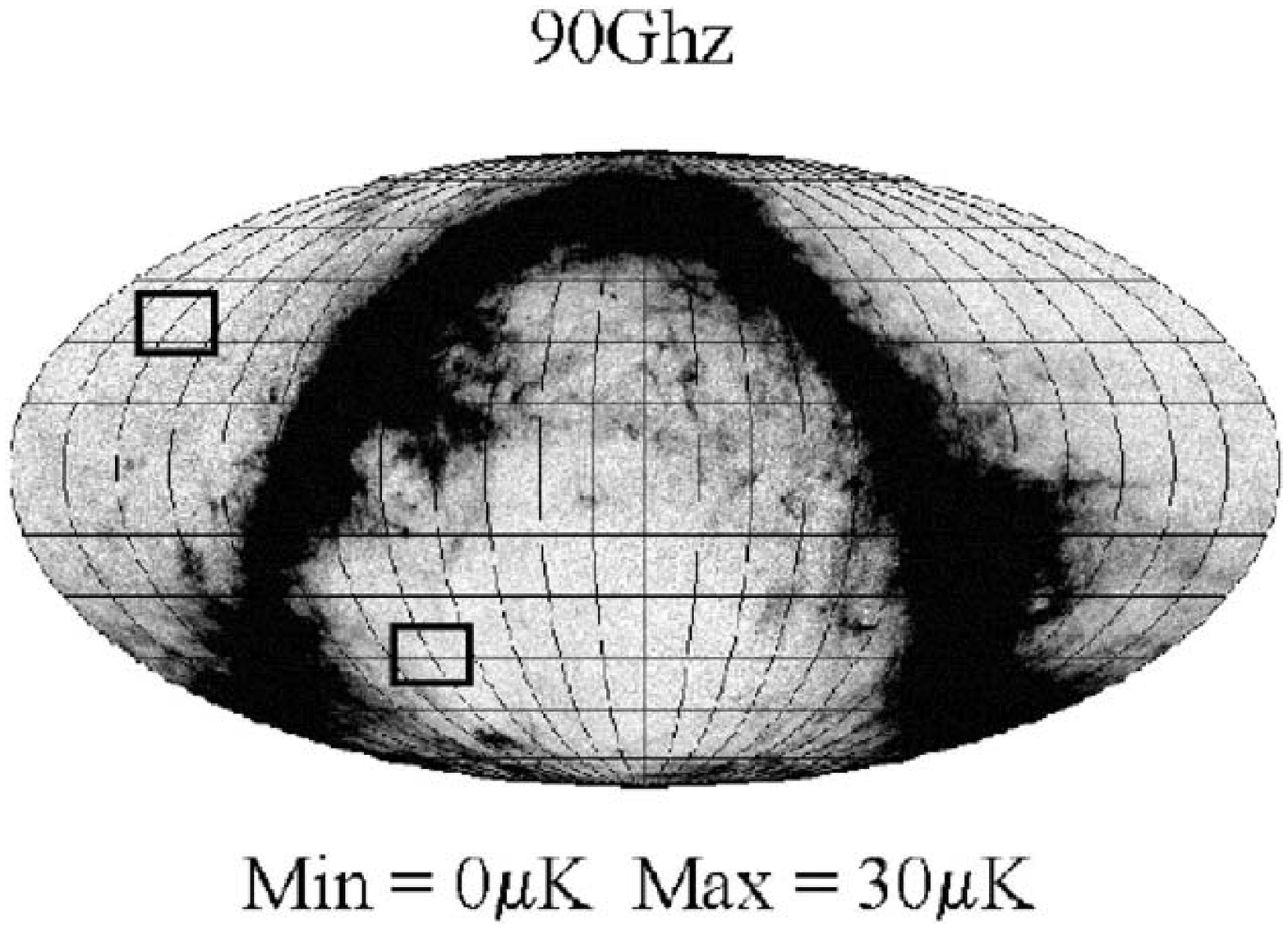}
\caption{The three foreground maps computed radio
\cite{1986A&AS...63..205R,1998MNRAS.297..977J} and far infra-red
\cite{1998ApJ...508...25H} data. The first two have been rescaled
up to 32 GHz to evaluate the synchrotron component, while the
last, related to the dust contribution, has been scaled down to 90
GHz. The two target patches have been superimposed.}
\label{multimaps}
\end{figure}

\noindent Both in the Northern and in the Southern hemisphere
interesting sky patches are present. The ideal regions, from the
CMB point of view, are the ones at high galactic latitude, far
from local cirrus, where the foreground should be minimal. In the
Southern sky, the patch observed by BOOMERanG
\cite{2000Natur.404..955D} ($\alpha=5^{h}, $
 $\delta=$-45$^\circ$) is the ideal target. Here the level of the
foregrounds at 32 GHz, where the expected main contribution comes
from synchrotron emission, can be evaluated from the
Jonas\cite{1998MNRAS.297..977J} map at 2.3 GHz scaling it with a
spectral index $\gamma=3$ (T$^{synch}\propto \nu^{-3}$) and with a
polarization fraction P$^{synch}/$T$^{synch}=0.1$. We can derive a
level of polarization $\Delta$P$<1\mu$K (making the synchrotron
contribution at 90 GHz even lower). For the 90 GHz channel, where
the dominant foreground contribution should come from dust, we
find $\Delta$P$<0.15\mu$K, starting from the DIRBE 240 $\mu$m map
\cite{1998ApJ...508...25H} and scaling the dust
temperature\cite{2000ApJ...530..133T} as T$^{dust}\propto
\nu^{2.7}/(e^{h\nu/kT_{D}}-1)$, using P$^{dust}/$T$^{dust}=0.05$.
In the Northern hemisphere for the region centered at
($\alpha=10^{h}, \delta=+35^\circ$) it is possible to derive
$\Delta P<0.3 \mu$K at 32 GHz and $\Delta P<0.15 \mu$K at 90 GHz.
For the latter we used again the DIRBE map, while for the former
the data from the Reich \cite{1986A&AS...63..205R} map at 1.4 GHz.
In Figure \ref{multimaps} the aforementioned maps
\cite{1986A&AS...63..205R,1998MNRAS.297..977J,1998ApJ...508...25H}
have been scaled as described for the evaluation of synchrotron
and dust contribution respectively at the BaR-SPOrt frequencies;
the selected patches both in the Northern and Southern sky are
shown.

\noindent From a logistical point of view, both patches are
effectively observable with Long Duration Balloon (LDB) flights.
The one in the Southern Sky is accessible from Antarctica, while
the one in the Northern hemisphere can be observed with a flight
launched from the available facilities both in Norway
\cite{2002apb..conf..243B} and in Sweden
\cite{2002apb..conf..239B}. Simulations are under development to
optimize the patch dimension, scanning speed and path for
maximizing the final full patch sensitivity. Expected raw
sensitivities are reported in Table \ref{maincharacteristics}.

\vspace{0.3cm}
\begin{table} [hbt]
\begin{center}
\begin{tabular}{cccccc}
\hline {Frequencies (GHz)} & {Bandwidth} & {Beamsize} &
{$\sigma_{1s}[{\rm mK}{\rm s}^{1/2}]$} &{$\sigma_{pix}^{2wfl[4wfl]}(\mu$K)}&{$\sigma_{prms}^{2wfl[4wfl]}(\mu$K)}\\
\hline 32 & 10\% & $0.4^\circ$ & 0.5 & 4.5 [3.2] & 3.0 [2.0] ({$2\,\sigma$ up. lim.})\\
\hline 90& 20\% & $0.2^\circ$ & 0.5 & 4.5 [3.2] & 0.7 [0.4]\\
\hline
\end{tabular}
\end{center}
\caption{BaR-SPOrt expected sensitivities: $\sigma_{1s}$ is the
instantaneous sensitivity, {$\sigma_{pix}^{2wfl[4wfl]}$} is the
pixel sensitivity considering a 2~[4] week flight and 100 pixel
patch; {$\sigma_{prms}^{2wfl[4wfl]}$} is the corresponding
sensitivity on P$_{rms}$. Due to the low P$_{rms}$ foreseen at 32
GHz ($2.1\,\mu$K against $4.0\,\mu$K at 90 GHz) the low frequency
channel is expected to provide upper limits.}
\label{maincharacteristics}
\end{table}

\section{The Instrument}

\begin{figure} [h]
\begin{center}
\includegraphics[width=5cm,angle=90]{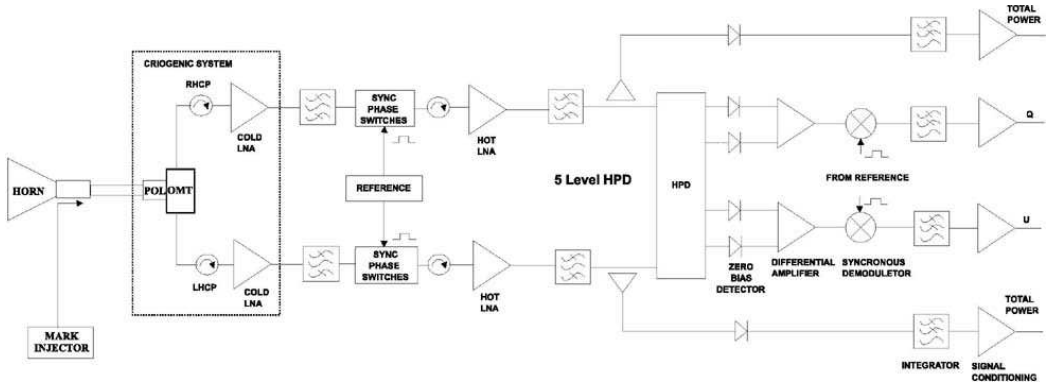} 
\caption{Block diagram of the BaR-SPOrt radiometers.}
\label{blockdiagram}
\end{center}
\end{figure}

\begin{figure} [h]
\begin{center}
\includegraphics[height=10cm]{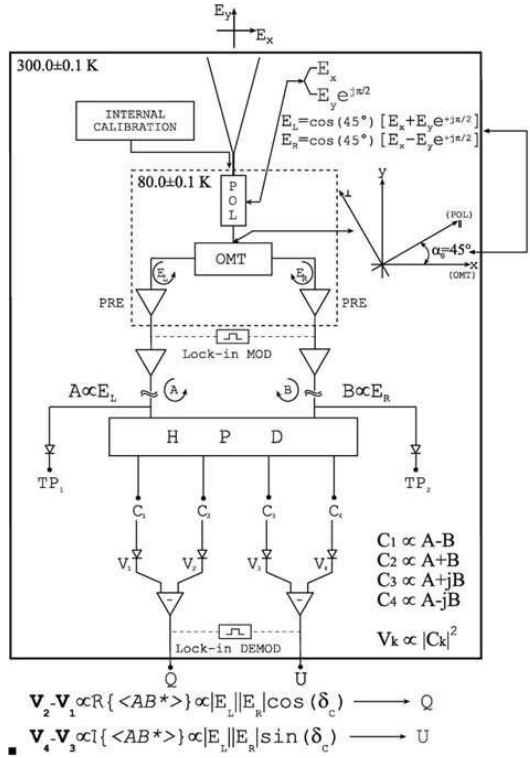}
\caption{Scheme of the radiometers with the propagation of the
fields collected by the feed horn.}
\label{operscheme}
\end{center}
\end{figure}

The BaR-SPOrt payloads house correlation microwave polarimeters
(see Figures \ref{blockdiagram} and \ref{operscheme}) for direct
measurement of the Q and U Stokes parameters with
HPBW=$0.4^{\circ}$ at 32 GHz and $0.2^{\circ}$ at 90 GHz. $1.8$~m
and $1.2$~m on axis Cassegrain configurations (for 32 and 90 GHz,
respectively) are adopted to meet the very stringent requirements
of extremely low spurious polarization (fractions of $\mu$K)
necessary for such measurements (for a general discussion see
Ref.~\citenum{2001NewA....6..173C}). The detection of signals as
low as those expected from CMB polarization implies the use of
extremely sensitive and stable radiometers. BaR-SPOrt has been
designed to minimize instrumental effects and to reduce 1/{\em f}
noise, thereby increasing the long term stability. Great care has
been taken to realize the antenna system keeping under control the
spurious polarization. The undesirable consequences of gain
fluctuations are greatly reduced by the correlation technique,
while residual instabilities are recovered using destriping data
analysis
\cite{2002apb..conf..193S,1998A&AS..127..555D,wrightastroph9612006,sbarraetalinprep,2000A&AS..142..499R},
which requires the radiometer to be stable only over a single scan
period (the scanning time of BaR-SPOrt is of the order of 1
minute).

The main instrumental characteristics are:
\begin{itemize}
\item direct amplification architecture: no down conversion to
avoid possible additional phase error; \item on axis low
cross-polarization optics providing HPBW of $0.4^\circ$
($0.2^\circ$) at 32 GHz (90 GHz); \item correlation unit based on
a custom design waveguide Hybrid Phase Discriminator (HPD),
capable of rejecting the unpolarized component better than 30 dB
\cite{2002ecmw..conf..150T,2002apb..conf..177P} and a phase
modulation (lock-in system) providing $> 60$ dB of total rejection
to the unpolarized component; \item custom design Orthomode
Transducer (OMT) with high isolation between channels ($> 60$ dB)
to limit contaminations from the unpolarized component; \item a
cryostat (see Figure \ref{chamber}) to cool down to $T < (80.0 \pm
0.1)$ K the Low Noise Amplifiers, the circulators, the polarizer
and the OMT by a closed loop cryocooler. The horn is kept, in the
present design, at $\approx $300 K, but might be cooled as well. A
thermal shield stabilized at temperature $T \cong (300.0 \pm 0.1)$
K, is adopted to increase the thermal stability of warm parts;
\item custom design internal calibrator to inject reference
polarized signals\cite{2002apb..conf..257B}.
\end{itemize}

The antenna system collects the incoming radiation and transforms,
by means of the polarizer, the linearly polarized components
($E_{x},E_{y}$) of the electric field into the circularly ones
($E_{R},E_{L}$) which are picked up by the Ortho-Mode
Transducer\cite{2002ecmw..conf..145M} (see Figures
\ref{operscheme}).

The lock-in detection is obtained by modulation ($0-\pi$ phase
shift) of one of the signals of the two chains just after the
first stage of amplification (LNA) and then by synchronous
detection in the back-end. Just before correlation, but inside the
Hybrid Phase Discriminator, a fraction of the signal is picked-up
and fed into two total power detectors to record the sky
temperature and for monitoring of the system temperature too. The
heart of the correlation unit is the custom design
HPD\cite{2002ecmw..conf..150T} that processes the signal in order
to have four outputs proportional to:
\begin{equation}
\vec {E_{R}} - \vec {E_{L}} \hspace{0.5cm} \vec {E_{R}} + \vec
{E_{L}} \hspace{0.5cm} \vec {E_{R}} + j\vec {E_{L}} \hspace{0.5cm}
\vec {E_{R}} - j\vec {E_{L}} \hspace{0.5cm}
\end{equation}
After square law detection the four HPD outputs are:
\begin{eqnarray}
V_{1} \propto [|\vec {E_{L}}|^2 + |\vec {E_{R}}|^2 -2 \Re({\vec
{E_{L}}\cdot \vec {E_{R}}}^{*})]\\  \nonumber V_{2} \propto [|\vec
{E_{L}}|^2 + |\vec {E_{R}}|^2 +2 \Re({\vec {E_{L}}\cdot \vec
{E_{R}}}^{*})]\\  \nonumber V_{3} \propto [|\vec {E_{L}}|^2 +
|\vec {E_{R}}|^2 +2 \Im({\vec {E_{L}}\cdot \vec {E_{R}}}^{*})]\\
\nonumber V_{4} \propto [|\vec {E_{L}}|^2 + |\vec {E_{R}}|^2 -2
\Im({\vec {E_{L}}\cdot \vec {E_{R}}}^{*})]       \nonumber
\end{eqnarray}
which are properly differentiated to get as final outputs the two
quantities:
\begin{eqnarray}
V_{2}-V_{1} \propto |\vec {E_{L}}||\vec {E_{R}}| \cos (\delta_{c}) \longrightarrow Q\\
V_{4}-V_{3} \propto |\vec {E_{L}}||\vec {E_{R}}| \sin (\delta_{c})
\longrightarrow U\nonumber
\end{eqnarray}
where $\delta_{c}$ is the phase delay between the two (L\&R)
circular components of the  electric field. After integration,
these provide time averaged values proportional to the Q and U
Stokes parameters.

\section{Preliminary Performance Evaluation of Some Custom Developed
Components of the 32 GHz Radiometer}

\begin{figure} [h]
\begin{center}
\includegraphics[width=5.2cm]{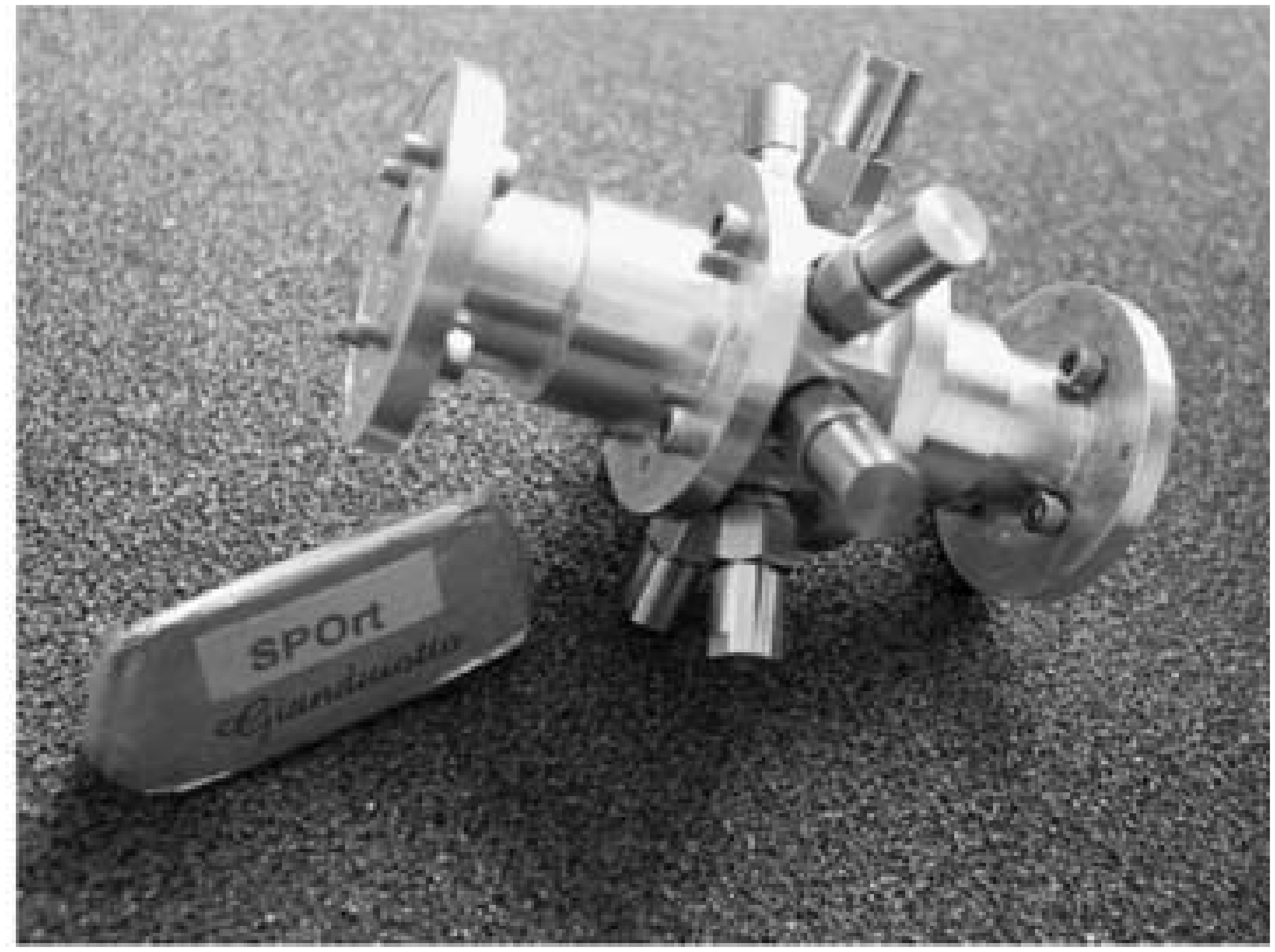}
\hspace{.1 cm}
\includegraphics[width=5.2cm]{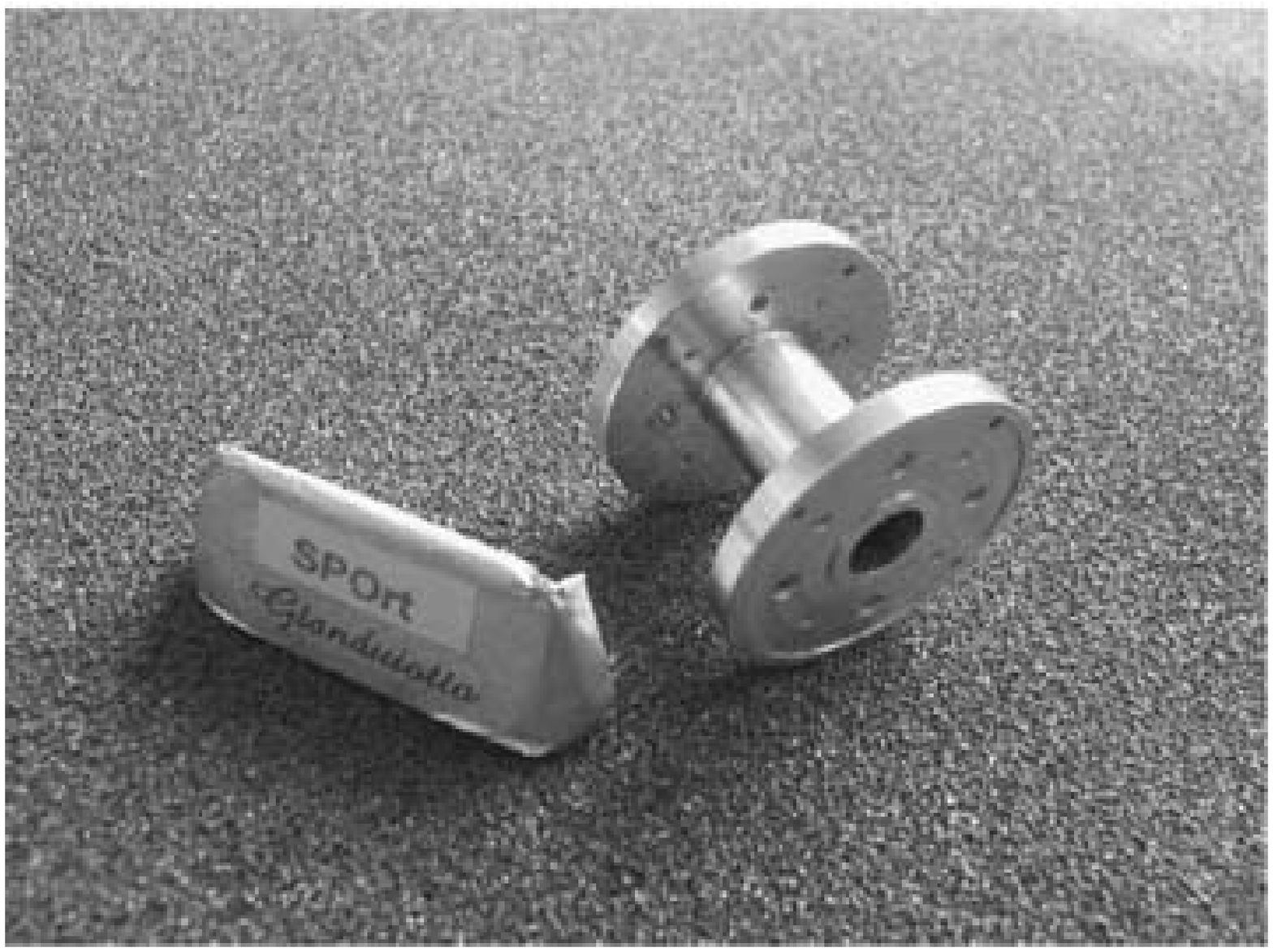}
\hspace{.1 cm}
\includegraphics[width=5.2cm,height=3.9cm]{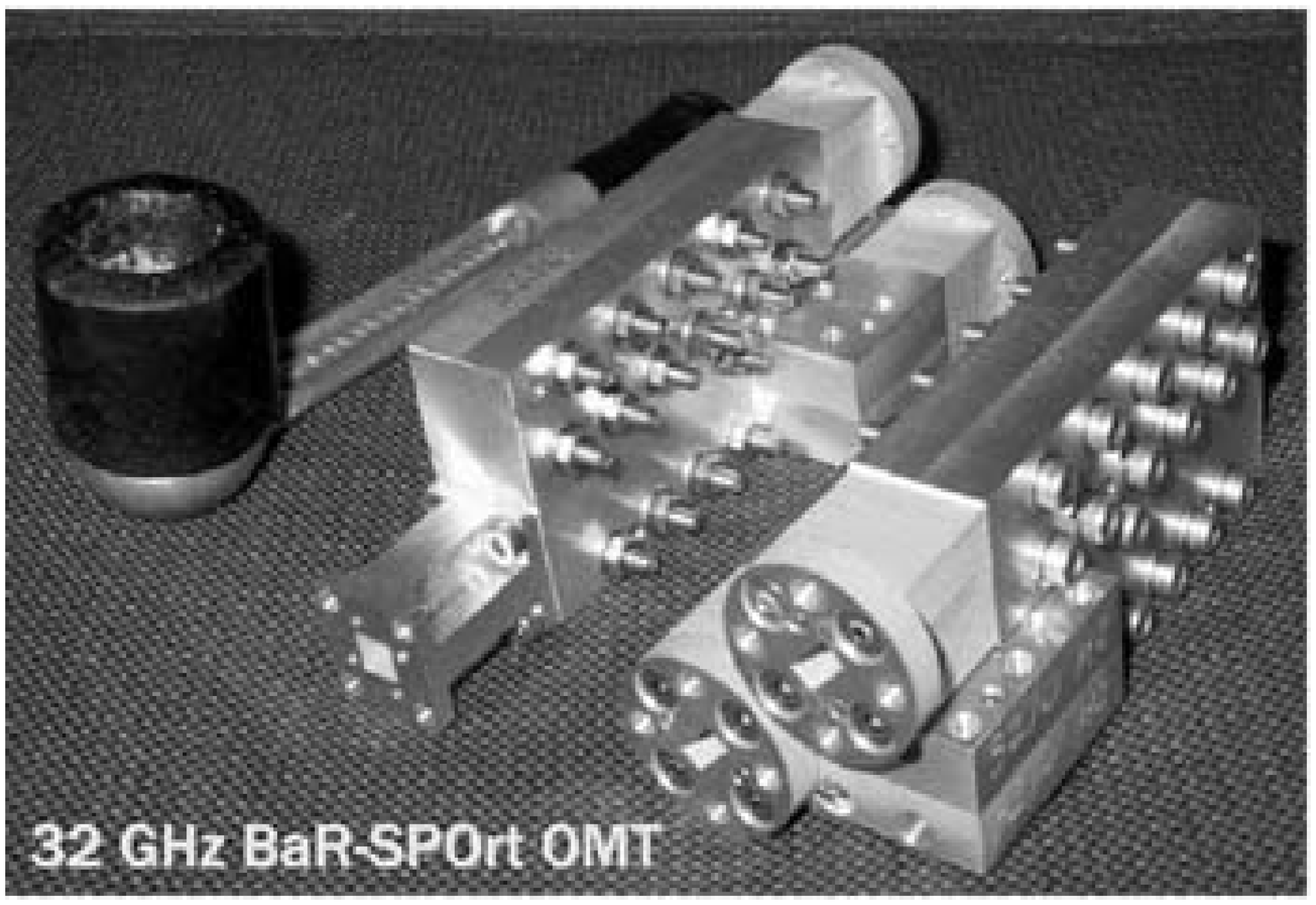}
\caption{{\bf Left image} shows the marker-injector. The device
has 8 points where to inject polarized reference signals. Only
three of them are fed, the remaining are for symmetry reasons. See
text for a more detailed description. {\bf Central image} shows
the Iris Polarizer realized in circular waveguide. {\bf Right
image} shows the high isolation OrthoMode Transducer.}
\label{photo_devices}
\end{center}
\end{figure}

\begin{figure} [h]
\begin{center}
\includegraphics[height=17cm,angle=90]{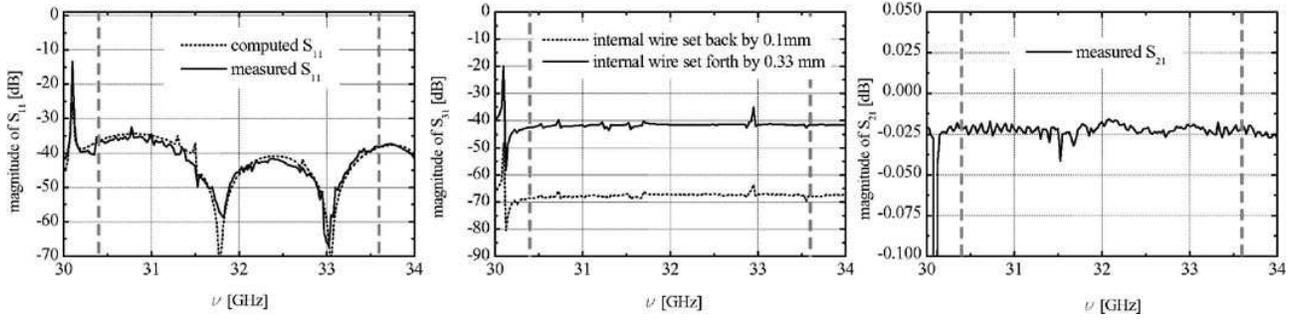}
\caption{{\em S} parameters for the {\em marker injector}: dotted
lines are for simulations while solid ones are for experimental
data from the real device. Vertical dashed lines show the
BaR-SPOrt bandwidth. {\bf Left panel} shows the S$_{11}$
parameter. {\bf Central panel} shows two different coupling
factors of reference polarized signals: adjusting the position of
the injector pins which feed the reference signals it is possible
to fine tune the coupling factor. {\bf Right panel} shows the
overall insertion loss of the device. The measured data are for an
aluminum device before silver plating.} \label{mark_injector}
\end{center}
\end{figure}

The tininess of the CMBP signal translates in very stringent
requirements of some critical components of the radiometer. These
are mainly the ones where the polarizations propagate
together\cite{2001NewA....6..173C,Carrettietalthisvolume}: in the
front-end we find, after the feed horn, the Reference Marker
Injector (the calibrator), the Polarizer and the OrthoMode
Transducer while in the back-end the Hybrid Phase Discriminator.
Here we report only about the devices in the front-end, the
performances of the HPD being already evaluated in
Ref.~\citenum{2002ecmw..conf..150T,2002apb..conf..177P}.
\newline \noindent The Reference Marker Injector is crucial in
experiments like BaR-SPOrt where external devices like wire grids
or diplexers to periodically feed the radiometer with known
polarization signals are difficult, if not impossible, to be used.
We have realized a marker injector\cite{2002apb..conf..257B}
(Figure \ref{photo_devices} {\bf left}) able to inject polarized
signals with three different polarization angles (nominally
$\frac{\pi}{8}, \frac{\pi}{8}\pm\frac{\pi}{4}$). The S parameters
of this device are reported in Figure \ref{mark_injector}. The
coupling factor can be adjusted simply changing the length of the
pins spilling the radiation (central panel). This is very
important both to reduce and control the amplitude of the mark,
produced by a noise generator, and not to degrade the polarization
status of the travelling sky signal. The last panel shows the
insertion loss of the device without silver coating: the very low
S$_{21}$ parameter ($\sim-0.025$~dB) will be even lower with the
foreseen silver plating.

\noindent With these reference signals periodically injected,
adopting the calibration technique described in
Ref.~\citenum{2002apb..conf..257B}, it is possible to fully
reconstruct the transfer matrixes of the instrument.

\begin{figure} [hbt]
\begin{center}
\includegraphics[height=12.5cm,angle=90]{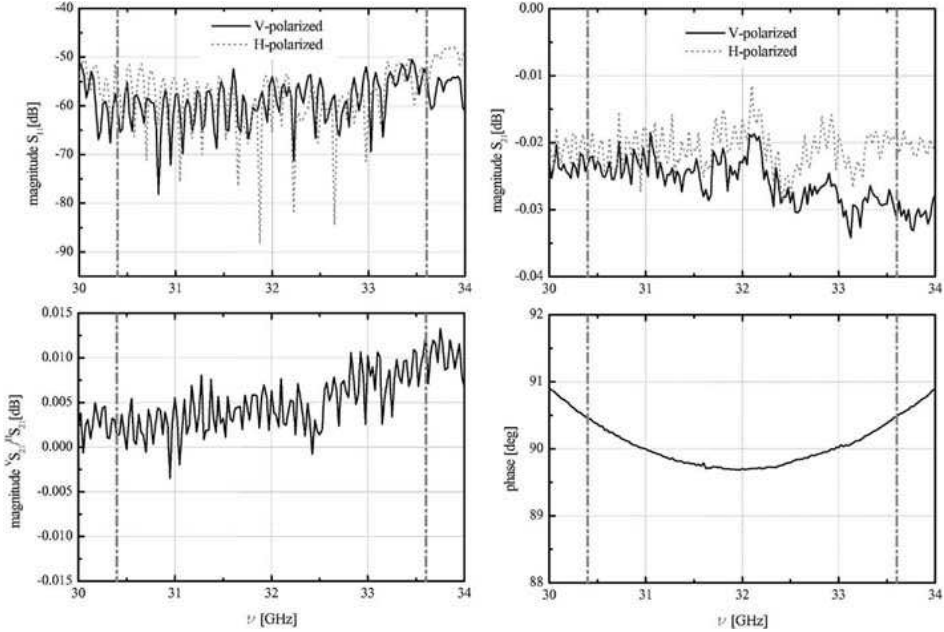}
\caption{The characteristic of the polarizer. {\bf Up left panel}
shows the reflection coefficient at the input port of the
polarizer. {\bf Up right panel} reports the transmission
coefficients (S$_{21}$) of the two perpendicular polarizations.
{\bf Down left panel} shows the difference between the two
S$_{21}$ parameters. {\bf Down right panel} is the phase
difference between the two polarizations. Vertical dashed lines
show the BaR-SPOrt bandwidth.} \label{polariser}
\end{center}
\end{figure}

\noindent After the marker injector and a waveguide thermal choke,
there is the cold part of the radiometer housing the polarizer,
the OMT and the LNAs. The polarizer, visible in the central panel
of Figure \ref{photo_devices}, is realized inserting some irises
in a circular waveguide. Great care has been taken to keep the
transmission coefficients for both polarizations extremely
similar: in the device realized their mean and maximum ratio are
$0.0047\ $dB and $0.01\ $dB respectively. Since a difference in
the S$_{21}$ parameters means an offset generation, their
equalizations was a key-point in the device design. The
$\frac{\pi}{2}$ phase difference between the polarizations is
constant too. As a matter of fact it is possible to show that a
not constant phase difference inside the integration bandwidth
translates into depolarization. The BaR-SPOrt custom polarizer
keeps this difference within $0.48^\circ$ (with a mean value of
$89.96^\circ$). The S$_{11}$ parameters are always below $-50\ $dB
all over the bandwidth (Figure \ref{polariser}).

\begin{figure} [h]
\begin{center}
\includegraphics[height=17cm,angle=90]{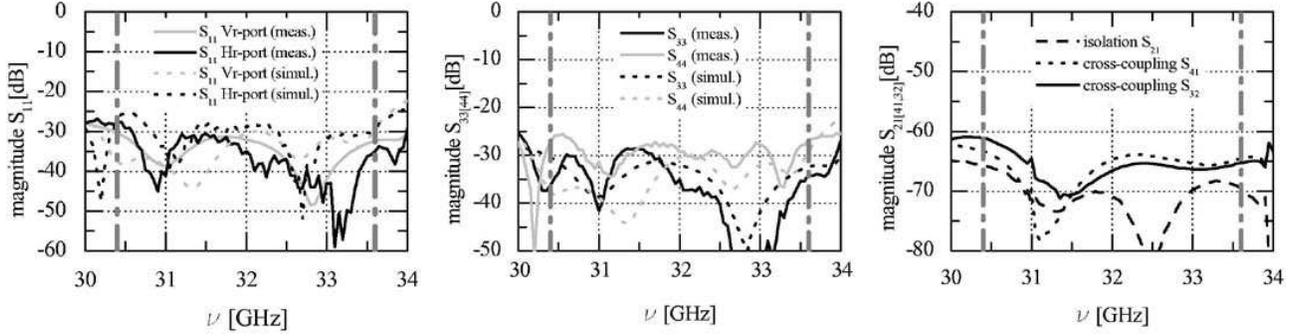}
\caption{{\em S} parameters for the {\em OMT}: dotted lines are
for simulations while solid ones are for experimental data from
the real device. Vertical dashed lines show the BaR-SPOrt
bandwidth. {\bf Left panel} shows the S$_{11}$ parameters of the
rectangular output ports. {\bf Central panel} shows the reflection
coefficients for the square input port. {\bf Right panel} shows
both the isolation and the cross-coupling: S$_{21}$ measures the
signal arriving in the horizontal port when the vertical one is
fed and the square input is closed with a matched load. S$_{41}$
measures the horizontal polarization contaminating the vertical
one and S$_{32}$ vice-versa.} \label{omt}
\end{center}
\end{figure}

The OMT is the device devoted to the polarization separation. Its
ability to correctly separate the vertical from the horizontal
polarization impacts on the final sensitivity of the
experiment\cite{2001NewA....6..173C}. The cross-talk and isolation
between the two polarizations measure the quality of this device.
In the BaR-SPOrt OMT we reached an extremely low polarization
contamination, both cross-talks and isolation being always well
below $-60\ $dB (Figure \ref{omt}).

\begin{figure} [htb]
\begin{center}
\includegraphics[width=8cm]{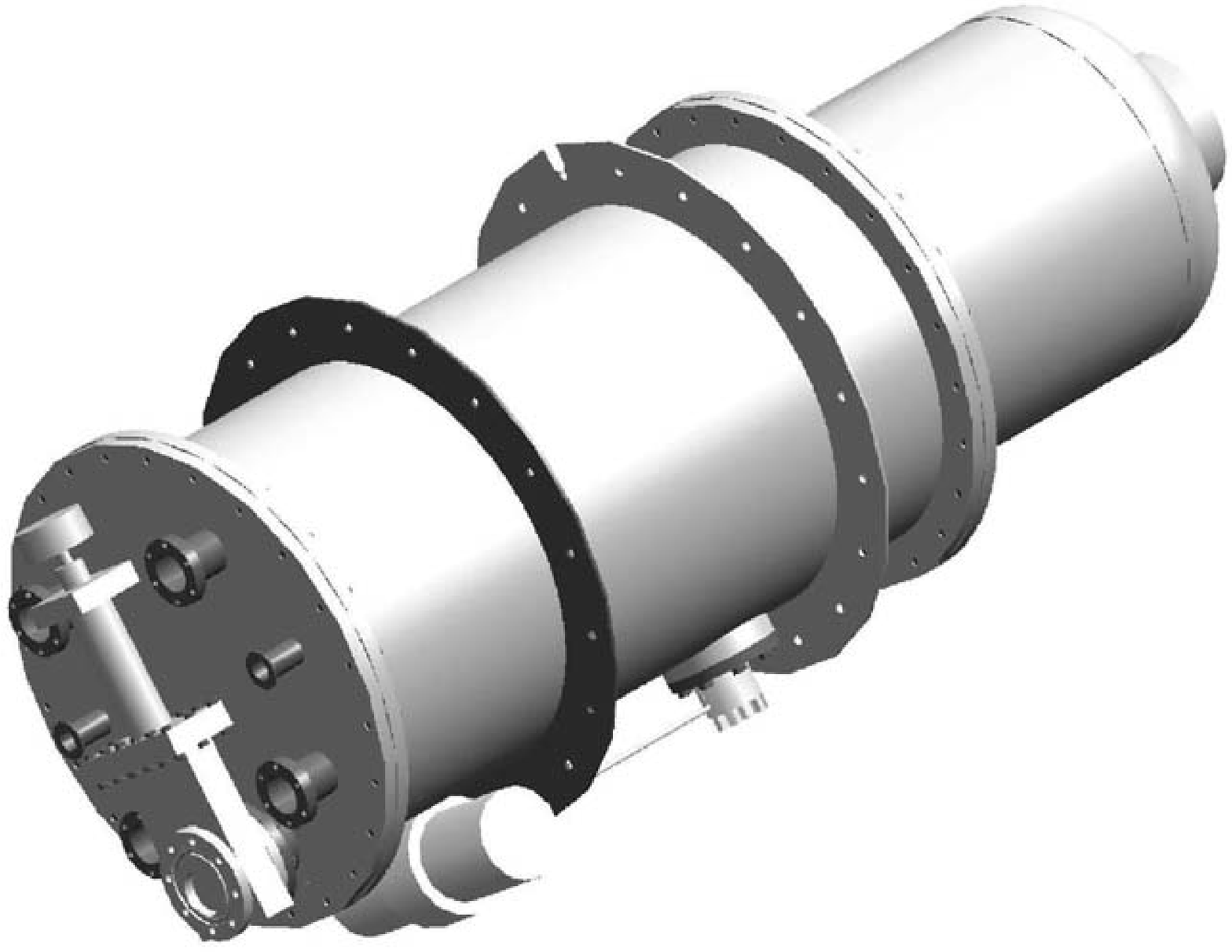}
\hspace{0.01 cm}
\includegraphics[width=8cm]{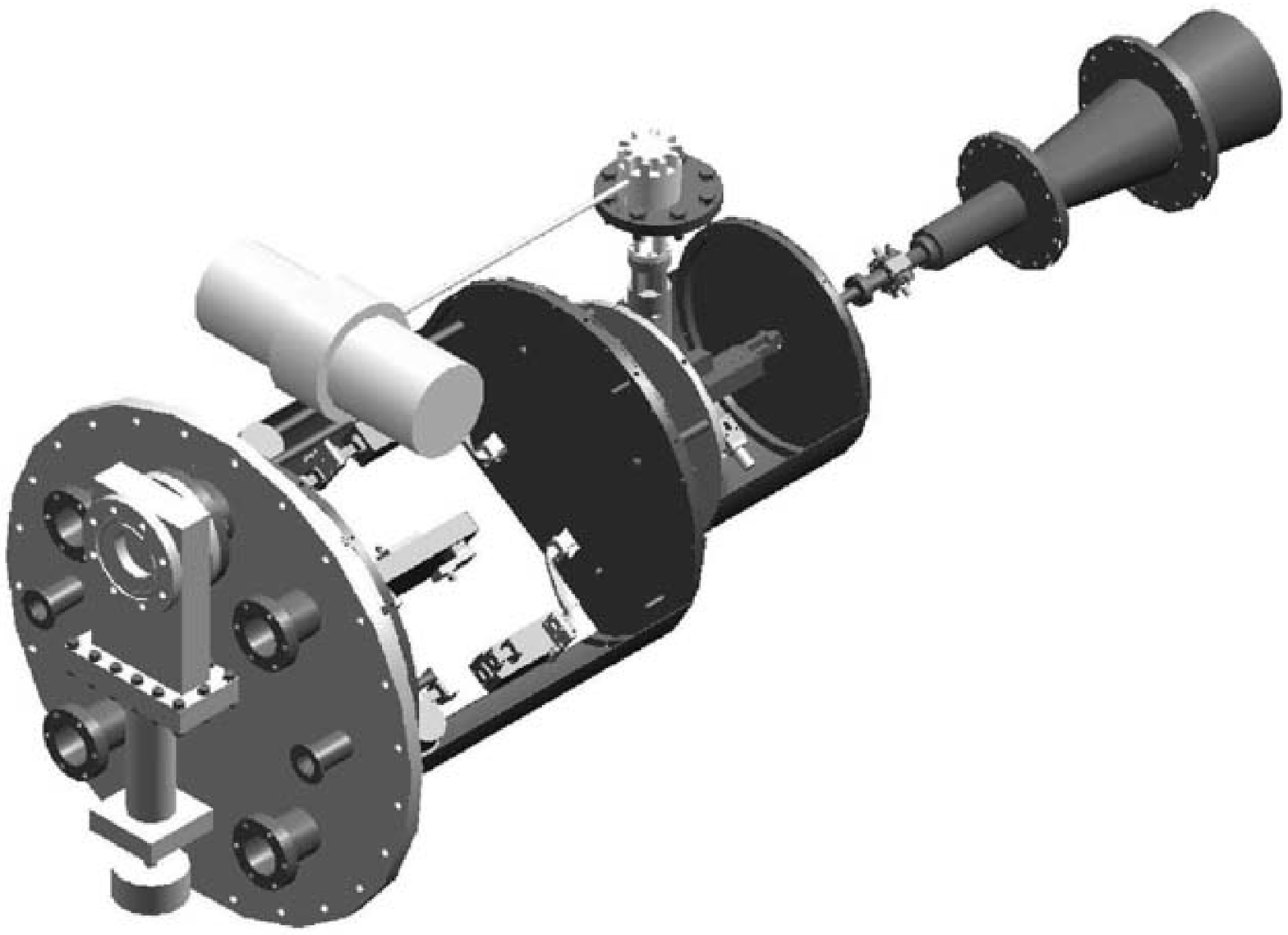}
\caption{The cryostat (left) housing the 32 GHz radiometer of
BaR-SPOrt with the closed loop cryocooler visible in the
foreground. The radiometer inside the cryostat (right).}
\label{chamber}
\end{center}
\end{figure}

\begin{figure} [htb]
\begin{center}
\includegraphics[width=7cm, angle=90]{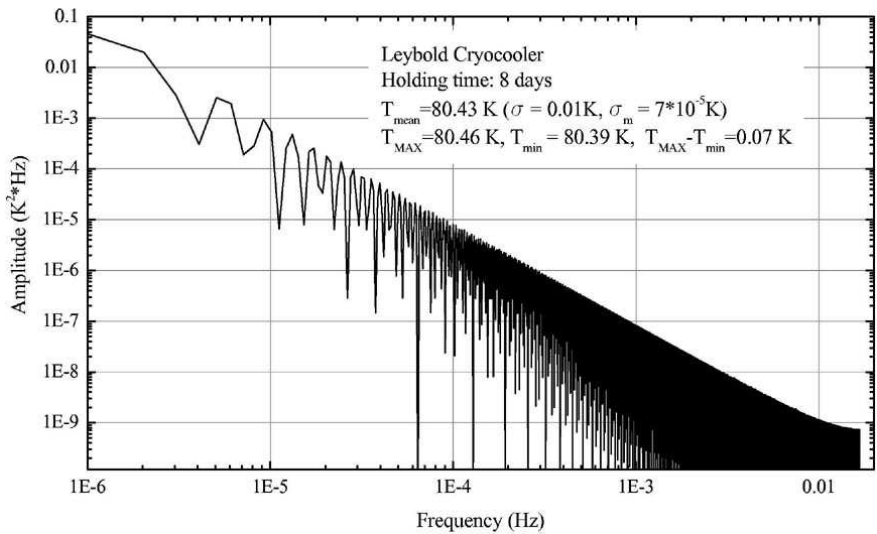}
\caption{Power Spectrum of the Leybold cryocooler computed for 8
days of dry run. The cooler has a very good stability thanks to
the PID algorithm governing the heat sink temperature.}
\label{cryo_test}
\end{center}
\end{figure}

Since the BaR-SPOrt performances strongly depend on the thermal
stability, particular care has been taken in the thermal design
(see Figure \ref{chamber}). The cooling system of BaR-SPOrt is
based on a mechanical Stirling cryocooler\footnote{Leybold model
Polar SC-7 COM} with closed loop control. Such cryogenerator can
cool the cold part of the radiometer down to 80 K with stability
better than 0.1 K. A laboratory dry run on the cooler lasted about
8 days, has shown a very good stability. Without any thermal
control over the environment (the laboratory) and any mass with
high thermal inertia anchored to the cold finger, the cooler was
extremely stable (see Figure \ref{cryo_test}), showing the high
effectiveness of the controller PID algorithm (for some more
preliminary results on the cooling system see Macculi {\em et
al.}\cite{2002apb..conf..275M}).

\section{Conclusions}
The BaR-SPOrt experiment is one of the first instruments with the
potentiality to measure, in its 90 GHz configuration, the CMB
polarization on sub-degree angular scale. When operated in the 32
GHz configuration, even if probably not enough sensitive for a
CMBP detection, it will permit to refine the existing upper limit
on the CMBP and will give data extremely important for the studies
of the polarized galactic foregrounds, in the light of the
BaR-SPOrt 90 GHz flight and the incoming CMB space missions (SPOrt
and, later, Planck). BaR-SPOrt represents also a very good
opportunity to test in operative conditions state of the art
technological solutions to be used in the SPOrt mission.

\acknowledgments
Authors wish to thank A.S.I. (Italian Space Agency) for the full
support to BaR-SPOrt and P.N.R.A. (National Project for Research
in Antarctica) for its relevant contribution. E.N.V. and M.V.S.
are grateful to C.N.R. (National Council for Research) for
supporting their collaboration within the C.N.R.-R.A.S (Russian
Academy of Science) agreement. We acknowledge use of CMBFAST and
HEALPix packages for performing our analysis.
\bibliography{SPIE_BAR-SPOrt}   

\begin{thebibliography}{10}

\bibitem{1982PLB....115...189S}
V.~A. {Rubakov}, M.~V. {Sazhin}, and A.~V. {Veryaskin}, ``{Graviton creation in
  the inflationary universe and the grand unification scale},'' {\em Physics
  Letters B} , 1982.

\bibitem{1990bmc..book.....D}
A.~D. {Dolgov}, M.~V. {Sazhin}, and Y.~B. {Zeldovich}, {\em {Basics of modern
  cosmology}}, Gif-sur-Yvette, France, Editions Frontieres, 1990, 251
  p.~Translation., 1990.

\bibitem{1996PhRvD..54.1332J}
G.~{Jungman}, M.~{Kamionkowski}, A.~{Kosowsky}, and D.~N. {Spergel},
  ``{Cosmological-parameter determination with microwave background maps},''
  {\em Physical Review D} {\bf 54}, pp.~1332--1344, July 1996.

\bibitem{1997ApJ...488....1Z}
M.~{Zaldarriaga}, D.~N. {Spergel}, and U.~{Seljak}, ``{Microwave Background
  Constraints on Cosmological Parameters},'' {\em Astrophysical Journal} {\bf
  488}, pp.~1+, Oct. 1997.

\bibitem{1999MNRAS.304...75E}
G.~{Efstathiou} and J.~R. {Bond}, ``{Cosmic confusion: degeneracies among
  cosmological parameters derived from measurements of microwave background
  anisotropies},'' {\em Monthly Notices of the Royal Astronomical Society} {\bf
  304}, pp.~75--97, Mar. 1999.

\bibitem{1965ApJ...142..419P}
A.~A. {Penzias} and R.~W. {Wilson}, ``{A Measurement of Excess Antenna
  Temperature at 4080 Mc/s.},'' {\em Astrophysical Journal} {\bf 142},
  pp.~419--421, July 1965.

\bibitem{1985ApJ...291L..23S}
G.~F. {Smoot}, G.~{de Amici}, S.~D. {Friedman}, C.~{Witebsky}, G.~{Sironi},
  G.~{Bonelli}, N.~{Mandolesi}, S.~{Cortiglioni}, G.~{Morigi}, and R.~B.
  {Partridge}, ``{Low-frequency measurements of the cosmic background radiation
  spectrum},'' {\em Astrophysical Journal Letters} {\bf 291}, pp.~L23--L27,
  Apr. 1985.

\bibitem{1996ApJ...473..576F}
D.~J. {Fixsen}, E.~S. {Cheng}, J.~M. {Gales}, J.~C. {Mather}, R.~A. {Shafer},
  and E.~L. {Wright}, ``{The Cosmic Microwave Background Spectrum from the Full
  COBE FIRAS Data Set},'' {\em Astrophysical Journal} {\bf 473}, pp.~576+, Dec.
  1996.

\bibitem{1999tkc..conf..165Z}
M.~{Zannoni}, G.~{Boella}, G.~{Bonelli}, F.~{Cavaliere}, M.~{Gervasi},
  A.~{Lagostina}, A.~{Passerini}, G.~{Sironi}, and A.~{Vaccari}, ``{TRIS
  Experiment: A Search for Spectral Distortions in the CMB Spectrum Close to 1
  GHz},'' in {\em AIP Conf. Proc. 476: 3K cosmology},  pp.~165+, 1999.

\bibitem{1996ApJ...464L...1B}
C.~L. {Bennett}, A.~J. {Banday}, K.~M. {Gorski}, G.~{Hinshaw}, P.~{Jackson},
  P.~{Keegstra}, A.~{Kogut}, G.~F. {Smoot}, D.~T. {Wilkinson}, and E.~L.
  {Wright}, ``{Four-Year COBE DMR Cosmic Microwave Background Observations:
  Maps and Basic Results},'' {\em Astrophysical Journal Letters} {\bf 464},
  pp.~L1--+, June 1996.

\bibitem{2000Natur.404..955D}
P.~{de Bernardis}, P.~A.~R. {Ade}, J.~J. {Bock}, J.~R. {Bond}, J.~{Borrill},
  A.~{Boscaleri}, K.~{Coble}, B.~P. {Crill}, G.~{De Gasperis}, P.~C. {Farese},
  P.~G. {Ferreira}, K.~{Ganga}, M.~{Giacometti}, E.~{Hivon}, V.~V. {Hristov},
  A.~{Iacoangeli}, A.~H. {Jaffe}, A.~E. {Lange}, L.~{Martinis}, S.~{Masi},
  P.~V. {Mason}, P.~D. {Mauskopf}, A.~{Melchiorri}, L.~{Miglio}, T.~{Montroy},
  C.~B. {Netterfield}, E.~{Pascale}, F.~{Piacentini}, D.~{Pogosyan},
  S.~{Prunet}, S.~{Rao}, G.~{Romeo}, J.~E. {Ruhl}, F.~{Scaramuzzi},
  D.~{Sforna}, and N.~{Vittorio}, ``{A flat Universe from high-resolution maps
  of the cosmic microwave background radiation},'' {\em Nature} {\bf 404},
  pp.~955--959, Apr. 2000.

\bibitem{2000ApJ...545L...5H}
S.~{Hanany}, P.~{Ade}, A.~{Balbi}, J.~{Bock}, J.~{Borrill}, A.~{Boscaleri},
  P.~{de Bernardis}, P.~G. {Ferreira}, V.~V. {Hristov}, A.~H. {Jaffe}, A.~E.
  {Lange}, A.~T. {Lee}, P.~D. {Mauskopf}, C.~B. {Netterfield}, S.~{Oh},
  E.~{Pascale}, B.~{Rabii}, P.~L. {Richards}, G.~F. {Smoot}, R.~{Stompor},
  C.~D. {Winant}, and J.~H.~P. {Wu}, ``{MAXIMA-1: A Measurement of the Cosmic
  Microwave Background Anisotropy on Angular Scales of 10'-5\&deg;},'' {\em
  Astrophysical Journal Letters} {\bf 545}, pp.~L5--L9, Dec. 2000.

\bibitem{1985SvAL...11..204S}
M.~V. {Sazhin} and V.~A. {Korolev}, ``{Polarization of the Unresolved-Source
  Microwave Background},'' {\em Soviet Astronomy Letters} {\bf 11}, pp.~204--+,
  July 1985.

\bibitem{1995ApL....32..105S}
M.~V. {Sazhin} and N.~{Benitez}, ``{Detecting gravitational waves via the
  Cosmic Microwave Background Polarization},'' {\em Astrophysical Letters and
  Communications} {\bf 32}, pp.~105--+, 1995.

\bibitem{2002apb..conf...109C}
E.~{Carretti}, M.~{Baralis}, G.~{Bernardi}, G.~{Boella}, S.~{Bonometto},
  M.~{Bruscoli}, S.~{Cecchini}, S.~{Cortiglioni}, R.~{Fabbri}, M.~{Gervasi},
  M.~{Gervasi}, C.~{Macculi}, J.~{Monari}, K.~{Ng}, L.~{Nicastro}, A.~{Orfei},
  O.~{Peverini}, S.~{Poppi}, V.~{Razin}, M.~{Sazhin}, C.~{Sbarra}, G.~{Sironi},
  I.~{Strukov}, R.~{Tascone}, M.~{Tucci}, E.~{Vinyajkin}, and M.~{Zannoni},
  ``{The SPOrt Experiment},'' in {\em AIP Conf. Proc. 609: Astrophysical
  Polarized Backgrounds},  pp.~109+, 2002.

\bibitem{1978PhRvD..17.1901C}
N.~{Caderni}, R.~{Fabbri}, B.~{Melchiorri}, F.~{Melchiorri}, and V.~{Natale},
  ``{Polarization of the microwave background radiation. I - Anisotropic
  cosmological expansion and evolution of the polarization states. II - an
  infrared survey of the sky},'' {\em Physical Review D} {\bf 17},
  pp.~1901--1918, Apr. 1978.

\bibitem{1979ApJ...232..341N}
G.~P. {Nanos}, ``{Polarization of the blackbody radiation at 3.2
  centimeters},'' {\em Astrophysical Journal} {\bf 232}, pp.~341--347, Sept.
  1979.

\bibitem{1981ApJ...245....1L}
P.~M. {Lubin} and G.~F. {Smoot}, ``{Polarization of the cosmic background
  radiation},'' {\em Astrophysical Journal} {\bf 245}, pp.~1--17, Apr. 1981.

\bibitem{1988Natur.331..146P}
R.~B. {Partridge}, J.~{Nowakowski}, and H.~M. {Martin}, ``{Linear polarized
  fluctuations in the cosmic microwave background},'' {\em Nature} {\bf 331},
  pp.~146+, Jan. 1988.

\bibitem{1993ApJ...419L..49W}
E.~J. {Wollack}, N.~C. {Jarosik}, C.~B. {Netterfield}, L.~A. {Page}, and
  D.~{Wilkinson}, ``{A Measurement of the Anisotropy in the Cosmic Microwave
  Background Radiation at Degree Angular Scales},'' {\em Astrophysical Journal
  Letters} {\bf 419}, pp.~L49--+, Dec. 1993.

\bibitem{1995ApJ...445L..69N}
C.~B. {Netterfield}, N.~{Jarosik}, L.~{Page}, D.~{Wilkinson}, and E.~{Wollack},
  ``{The anisotropy in the cosmic microwave background at degree angular
  scales},'' {\em Astrophysical Journal Letters} {\bf 445}, pp.~L69--L72, June
  1995.

\bibitem{1997NewA....3....1S}
G.~{Sironi}, G.~{Boella}, G.~{Bonelli}, L.~{Brunetti}, F.~{Cavaliere},
  M.~{Gervasi}, G.~{Giardino}, and A.~{Passerini}, ``{A 33 GHz polarimeter for
  observations of the Cosmic Microwave Background},'' {\em New Astronomy} {\bf
  3}, pp.~1--13, Nov. 1997.

\bibitem{2000MNRAS.315..808S}
R.~{Subrahmanyan}, M.~J. {Kesteven}, R.~D. {Ekers}, M.~{Sinclair}, and
  J.~{Silk}, ``{An Australia Telescope survey for CMB anisotropies},'' {\em
  Monthly Notices of the Royal Astronomical Society} {\bf 315}, pp.~808--822,
  July 2000.

\bibitem{2002ApJ...573L..73H}
M.~M. {Hedman}, D.~{Barkats}, J.~O. {Gundersen}, J.~J. {McMahon}, S.~T.
  {Staggs}, and B.~{Winstein}, ``{New Limits on the Polarized Anisotropy of the
  Cosmic Microwave Background at Subdegree Angular Scales},'' {\em
  Astrophysical Journal Letters} {\bf 573}, pp.~L73--LL76, July 2002.

\bibitem{2001ApJ...560L...1K}
B.~G. {Keating}, C.~W. {O'Dell}, A.~{de Oliveira-Costa}, S.~{Klawikowski},
  N.~{Stebor}, L.~{Piccirillo}, M.~{Tegmark}, and P.~T. {Timbie}, ``{A Limit on
  the Large Angular Scale Polarization of the Cosmic Microwave Background},''
  {\em Astrophysical Journal Letters} {\bf 560}, pp.~L1--L4, Oct. 2001.

\bibitem{2001NewA....6..173C}
E.~{Carretti}, R.~{Tascone}, S.~{Cortiglioni}, J.~{Monari}, and M.~{Orsini},
  ``{Limits due to instrumental polarisation in CMB experiments at microwave
  wavelengths},'' {\em New Astronomy} {\bf 6}, pp.~173--187, May 2001.

\bibitem{1998MNRAS.297..977J}
J.~L. {Jonas}, E.~E. {Baart}, and G.~D. {Nicolson}, ``{The Rhodes/HartRAO
  2326-MHz radio continuum survey},'' {\em Monthly Notices of the Royal
  Astronomical Society} {\bf 297}, pp.~977--989, July 1998.

\bibitem{1998ApJ...508...25H}
M.~G. {Hauser}, R.~G. {Arendt}, T.~{Kelsall}, E.~{Dwek}, N.~{Odegard}, J.~L.
  {Weiland}, H.~T. {Freudenreich}, W.~T. {Reach}, R.~F. {Silverberg}, S.~H.
  {Moseley}, Y.~C. {Pei}, P.~{Lubin}, J.~C. {Mather}, R.~A. {Shafer}, G.~F.
  {Smoot}, R.~{Weiss}, D.~T. {Wilkinson}, and E.~L. {Wright}, ``{The COBE
  Diffuse Infrared Background Experiment Search for the Cosmic Infrared
  Background. I. Limits and Detections},'' {\em Astrophysical Journal} {\bf
  508}, pp.~25--43, Nov. 1998.

\bibitem{2000ApJ...530..133T}
M.~{Tegmark}, D.~J. {Eisenstein}, W.~{Hu}, and A.~{de Oliveira-Costa},
  ``{Foregrounds and Forecasts for the Cosmic Microwave Background},'' {\em
  Astrophysical Journal} {\bf 530}, pp.~133--165, Feb. 2000.

\bibitem{1986A&AS...63..205R}
P.~{Reich} and W.~{Reich}, ``{A radio continuum survey of the northern sky at
  1420 MHz. II},'' {\em Astronomy and Astrophysics Supplement} {\bf 63},
  pp.~205--288, Feb. 1986.

\bibitem{2002apb..conf..243B}
K.~{B\o en}, ``{Scientific Ballons from Svalbard},'' in {\em AIP Conf. Proc.
  609: Astrophysical Polarized Backgrounds},  pp.~243+, 2002.

\bibitem{2002apb..conf..239B}
P.~{Baldemar} and O.~{Widell}, ``{The Esrange Facility in Northern Sweden--Your
  Partner for Successful Aerospace Operations},'' in {\em AIP Conf. Proc. 609:
  Astrophysical Polarized Backgrounds},  pp.~239+, 2002.

\bibitem{2002apb..conf..193S}
C.~{Sbarra}, E.~{Carretti}, S.~{Cortiglioni}, M.~{Zannoni}, R.~{Fabbri},
  C.~{Macculi}, and M.~{Tucci}, ``{A Destriping Technique for SPOrt
  Polarization Data},'' in {\em AIP Conf. Proc. 609: Astrophysical Polarized
  Backgrounds},  pp.~193+, 2002.

\bibitem{1998A&AS..127..555D}
J.~{Delabrouille}, ``{Analysis of the accuracy of a destriping method for
  future cosmic microwave background mapping with the PLANCK SURVEYOR
  satellite},'' {\em Astronomy and Astrophysics Supplement} {\bf 127},
  pp.~555--567, Feb. 1998.

\bibitem{wrightastroph9612006}
E.~L. {Wright}, ``{Scanning and Mapping Strategies for CMB Experiments. Paper
  presented at the IAS CMB Data Analysis Workshop in Princeton on 22 Nov 96},''
  in {\em \itshape{astro-ph}/9612006},  1996.

\bibitem{sbarraetalinprep}
C.~{Sbarra} and et~al., ``{in preparation},''

\bibitem{2000A&AS..142..499R}
B.~{Revenu}, A.~{Kim}, R.~{Ansari}, F.~{Couchot}, J.~{Delabrouille}, and
  J.~{Kaplan}, ``{Destriping of polarized data in a CMB mission with a circular
  scanning strategy},'' {\em Astronomy and Astrophysics Supplement} {\bf 142},
  pp.~499--509, Mar. 2000.

\bibitem{2002ecmw..conf..150T}
R.~{Tascone}, D.~{Trinchero}, M.~{Baralis}, O.~{Peverini}, A.~{Olivieri},
  E.~{Carretti}, and S.~{Cortiglioni}, ``{Millimeter Wave Passive Devices for
  Measurements of the Polarized Sky Emission},'' in {\em AIP Conf. Proc. 616:
  Experimental Cosmology at Millimetre Wavelengths},  pp.~150+, 2002.

\bibitem{2002apb..conf..177P}
O.~{Peverini}, M.~{Baralis}, R.~{Tascone}, D.~{Trinchero}, A.~{Olivieri},
  E.~{Carretti}, and S.~{Cortiglioni}, ``{Millimeter Wave Passive Components
  for Polarization Measurements},'' in {\em AIP Conf. Proc. 609: Astrophysical
  Polarized Backgrounds},  pp.~177+, 2002.

\bibitem{2002apb..conf..257B}
M.~{Baralis}, O.~{Peverini}, R.~{Tascone}, D.~{Trinchero}, V.~{Niculae},
  A.~{Olivieri}, E.~{Carretti}, S.~{Cortiglioni}, C.~{Macculi}, J.~{Monari},
  A.~{Orfei}, G.~{Sironi}, and M.~{Zannoni}, ``{Calibration Techniques and
  Devices for Correlation Radiometers Used in Polarization Measurements},'' in
  {\em AIP Conf. Proc. 609: Astrophysical Polarized Backgrounds},  pp.~257+,
  2002.

\bibitem{2002ecmw..conf..145M}
C.~{Macculi}, G.~{Bernardi}, E.~{Carretti}, S.~{Cecchini}, S.~M.~E.
  {Cortiglioni}, C.~{Sbarra}, V.~G., J.~{Monari}, S.~{Poppi}, G.~{Boella},
  S.~{Bonometto}, M.~{Gervasi}, G.~{Sironi}, M.~{Tucci}, M.~{Zannoni},
  M.~{Baralis}, O.~{Peverini}, R.~{Tascone}, R.~{Fabbri}, V.~{Natale},
  M.~{Bruscoli}, A.~{Boscaleri}, E.~{Pascale}, and L.~{Nicastro}, ``{BaR-SPOrt:
  a technical overview},'' in {\em AIP Conf. Proc. 616: Experimental Cosmology
  at Millimetre Wavelengths},  pp.~145+, 2002.

\bibitem{Carrettietalthisvolume}
E.~{Carretti} and et~al., ``{SPOrt: an Experiment Aimed at Measuring the Large
  Scale Cosmic Microwave Background Polarization},'' {\em this volume} .

\bibitem{2002apb..conf..275M}
C.~{Macculi} and M.~{Zannoni}, ``{Thermal Design and Preliminary Performance
  Evaluation of the Cooling System for BaR-SPOrt},'' in {\em AIP Conf. Proc.
  609: Astrophysical Polarized Backgrounds},  pp.~275+, 2002.

\end{thebibliography}
\bibliographystyle{spiebib}   

\end{document}